# Multiple boundary states in bilayer and decorated Su-Schrieffer-Heeger-like models

Shengqun Guo,[1] Jinke Huang,[1] Ruimin Huang,[1] Fengjiang Zhuang,[1] Zhili Lin,[1] and Weibin Qiu [1,2]*

[1] *College of Information Science and Engineering, Huaqiao University, Xiamen 361021, China*
[2] *Fujian Science & Technology Innovation Laboratory for Optoelectronic Information of China, Fuzhou 350108, China*

Topological boundary states have attracted widespread fascination due to their series of intriguing properties. In this paper, we investigate the multiple boundary states within the two kinds of extended Su-Schrieffer-Heeger (SSH) models. The coexistence of boundary states that exist both in the bulk and band gaps is realized based on the bilayer SSH-like model, which consists of two conventional square-root SSH models that are directly coupled. We further show the square-root topology within the decorated SSH-like model, which supports multiple boundary states that could be embedded into the bulk continuum by tuning the hopping parameters. In addition, the connection between the decorated SSH-like model and its effectively decomposed counterparts is revealed. Our results broaden insight into the multiple boundary states and open up an exciting avenue for the future exploration of square-root topology.

*Introduction.* Topological states of matter have arisen as an intriguing topic in the community of condensed matter physics[1-4], owing to their notable characteristic of topological boundary states that are not sensitive to local perturbations. Significant efforts have been devoted to the quest for novelty topological states. Therein, the celebrated SSH chain holds a crucial position in describing topological physics[5,6]. Various intriguing physics have been explored in the SSH chain and its extension versions, including but not limited to the two-dimensional weak topological insulators (TIs)[7,8], higher-order TIs[9,10], and twofold topological phase transitions induced by long-range hoppings[11].

Recently, the concept of square-root topology provided the scheme for constructing a new class of unconventional TIs[12], whose topological properties can be traced back to the parent Hamiltonian with the square-root operation. The corresponding tight-binding models can be obtained by inserting the additional sites. Such square-root TIs support the finite energy boundary states. Thus far, the square-root topology has been actively explored in first- and higher-order TIs[13-26], and several studies have reported that the concept can be extended to the quartic-root topology to further double the band gaps[27-31], with the quartic-root TIs generally exhibiting four groups of topological boundary states located in different gaps under the open boundary condition (OBC).

---

*[*]wbqiu@hqu.edu.cn

Furthermore, topological boundary states can be embedded in the bulk, which behave as the bound states in the continuum[32,33]. The typical case for unveiling this phenomenon is the two-dimensional SSH model[34-39], with support for the zero-dimensional corner states. Later research observed that the boundary states which emerge in the bulk spectrum can be realized in the trimerized lattice with proper loss settings[40]. Moreover, this phenomenon of topological boundary states is also studied in the mirror-stacking approach[41,42] and insight-related subspace-induced bound states in the continuum[43-46]. In this paper, we investigate the multiple boundary states based on the bilayer and decorated SSH-like models. We show that the bilayer SSH-like model could host the topological boundary states that emerge in the bulk and band gaps simultaneously, which are called in-bulk and in-gap boundary states in this work, respectively. Furthermore, we construct a decorated SSH-like model that hosts square-root topology and supports the four groups of boundary states. The squared Hamiltonian of this model correspond to the direct sum of the Hamiltonian of a bilayer SSH model and a residual Hamiltonian. By tuning the hopping parameters, the multiple boundary states can be shifted to the bulk, forming in-bulk boundary states. Interestingly, we also find that the associated Hamiltonian of this model could be decoupled into two independent subspaces, which correspond to the two types of square-root versions of the SSH chain. Our work brings about new possibilities for further exploring multiple boundary states and square-root topology.

*Bilayer SSH-like models.* As depicted in Fig. 1(a), we start by considering the bilayer SSH-like model, which consists of two identical square-root SSH models that are coupled directly, with the square-root SSH model that contains four sites per unit cell being the focus of research on square-root TIs[14,25-27]. The Hamiltonian of the bilayer SSH-like model in momentum space can be expressed as

$$H_1 = \begin{pmatrix} h_1 & h_2 \\ h_2 & h_1 \end{pmatrix}, \quad (1)$$

with the $4\times 4$ matrices

$$h_1 = \begin{pmatrix} 0 & 0 & t_a & t_b e^{-ik} \\ 0 & 0 & t_a & t_b \\ t_a & t_a & 0 & 0 \\ t_b e^{ik} & t_b & 0 & 0 \end{pmatrix}, \quad (2)$$

and

$$h_2 = \begin{pmatrix} t_c & 0 & 0 & 0 \\ 0 & t_c & 0 & 0 \\ 0 & 0 & t_c & 0 \\ 0 & 0 & 0 & t_c \end{pmatrix}, \quad (3)$$

where $k$ is the Bloch wave vector, $t_a$ and $t_b$ represent the intralayer hopping parameters, and $t_c$ represent the interlayer hopping parameters. After similarity transformation $S^{-1}H_1 S = \tilde{H}_1$, where $S = \begin{pmatrix} I_{4\times 4} & I_{4\times 4} \\ I_{4\times 4} & -I_{4\times 4} \end{pmatrix}$ ($I_{n\times n}$ is the $n\times n$ identity matrix), one can obtain

$$\tilde{H}_1 = \begin{pmatrix} \beta_1 & O_{4\times 4} \\ O_{4\times 4} & \beta_2 \end{pmatrix}, \quad (4)$$

where $\beta_1 = h_1 + h_2$, $\beta_2 = h_1 - h_2$, and $O_{4\times 4}$ is the zero matrix. Here, $\beta_1$ and $\beta_2$ correspond to the Hamiltonian of the square-root SSH models, with energy shift $t_c$ and $-t_c$, respectively. It should be noted that the bilayer SSH-like model preserves mirror symmetry along z-direction, with $[M_{z,1}, H_1] = 0$, where $M_{z,1} = \begin{pmatrix} O_{4\times 4} & I_{4\times 4} \\ I_{4\times 4} & O_{4\times 4} \end{pmatrix}$. By taking similarity transformation $S^{-1}M_{z,1}S = \tilde{M}_{z,1}$, the mirror operator is diagonalized into $\tilde{M}_{z,1} = \begin{pmatrix} I_{4\times 4} & O_{4\times 4} \\ O_{4\times 4} & -I_{4\times 4} \end{pmatrix}$, indicating that the subspace $\beta_1$ ($\beta_2$) corresponds to the even (odd) subspace according to mirror parity[41]. Figure. 1(b) shows the relationship between the bilayer SSH-like model and its effective decomposition counterparts. The nontrivial topological property of the square-root SSH models is exhibited for the case of $t_a < t_b$ [14,25-27], which leads to the presence of boundary states located in dual band gaps under the OBC. To satisfy the topological nontrival subspace, the following intralayer coupling parameter is fixed to $t_a = 1$, $t_b = 3$. The representative band structures with $t_c = 5$ are shown in Fig. 1 (c)–(d), with the bulk bands of the bilayer SSH-like model [Fig. 1(c)] decoupled into band structures corresponding to $\beta_1$ and $\beta_2$ [Fig. 1(d)] according to their mirror parities, which agree with the analyses above.

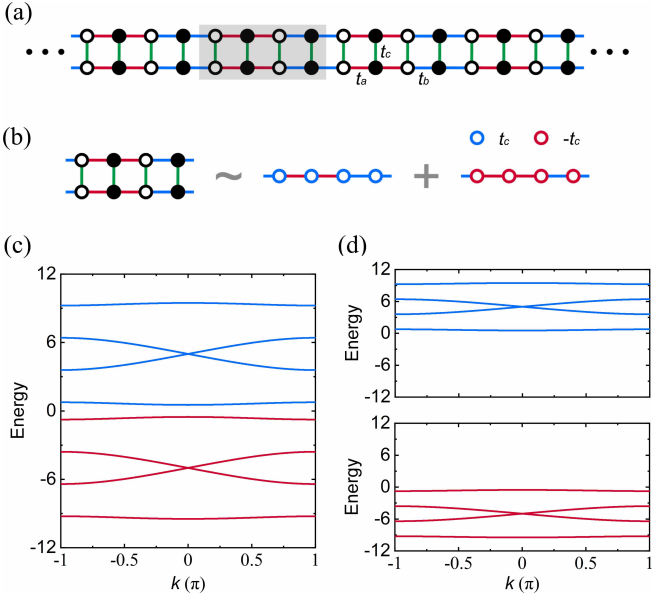

FIG. 1. (a) Schematic of the bilayer SSH-like model. (b) The relationship between the bilayer SSH-like model and its effective decomposition counterparts. (c)-(d) band structure of the (c) bilayer SSH-like model and (d) effective decomposition counterparts.

The corresponding finite-size structure with 10 unit cells is shown in Fig. 2(a). To reveal the existence of multiple boundary states, Fig. 2(b) exhibit the eigenvalues as a function of $t_c / t_b$ ($t_a = 1$, $t_b = 3$). One can observe that the boundary states can be embedded into the bulk by adjusting the interlayer coupling $t_c$, thus enabling the coexistence of in-bulk boundary states and in-gap boundary states. Furthermore, we chose $t_c = 1.5$ to further study the phenomena of topological boundary states, the corresponding energy spectrum in this case is shown in Fig. 2(c), where the red and blue dots represent the boundary and bulk states, respectively. As expected, two groups of boundary states are embedded in the bulk, while the other groups of boundary states emerge in the band gaps, manifested as the coexistence of in-bulk and in-gap boundary states. To demonstrate the localized features of the boundary states, we plot the distributions of one of the bulk states (state index $n = 6$) and the representative boundary states ($n = 10, 29$) that emerge in the band gap and bulk continuums, respectively. As shown in Fig. 2(d), one can see that the boundary states are indeed localized at the end region.

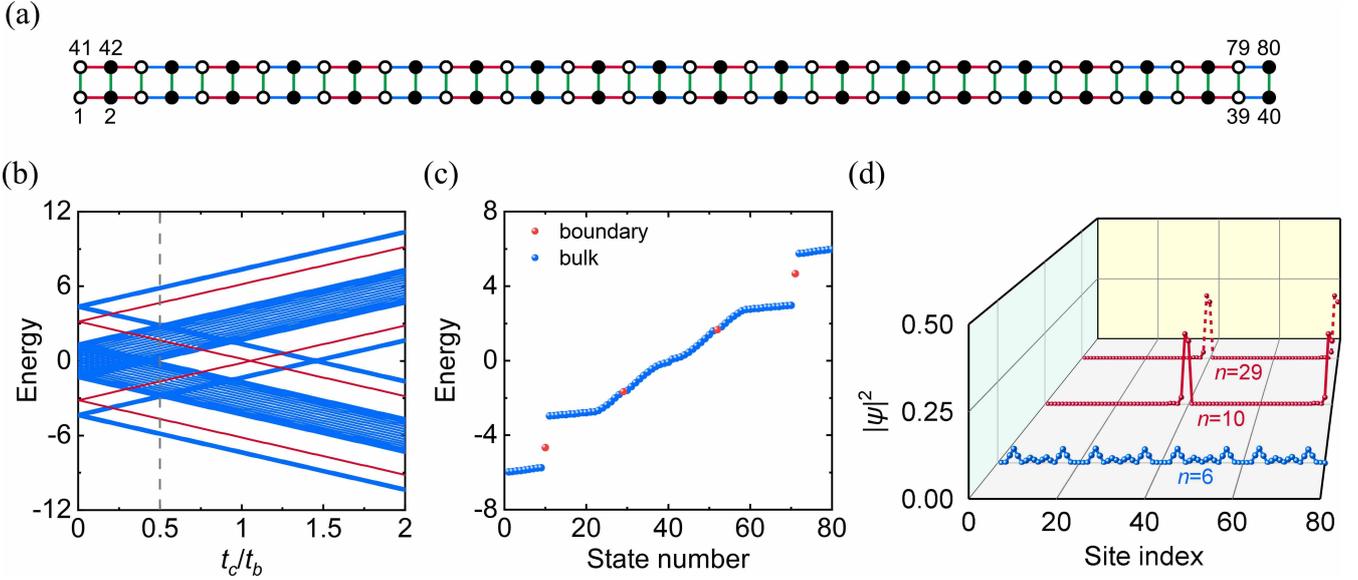

FIG. 2. (a) Schematic of the finite-size bilayer SSH-like model. (b) Energy spectrum for the finite-size bilayer SSH-like model with different $t_c$, where $t_a = 1$, $t_b = 3$ are fixed. (c) Energy spectrum at $t_c = 1.5$ [gray dotted line in (b)]. (d) The probability distributions of the bulk and boundary states.

The in-bulk boundary states in the finite-size bilayer SSH-like model are emerge through the suitable adjustment of interlayer coupling to make the boundary states move into the bulk continuums with different parities. The different parities can limit the hybridization between the topological boundary states and the bulk continuums[41]. Here, we introduce two types of weak perturbation by adding the next-nearest-neighbor (NNN) interlayer couplings to examine the robustness of the in-bulk boundary states. The NNN couplings are set as $t_d$ for convenience. When the NNN interlayer couplings that respects mirror symmetry is introduced [the inset of Fig. 3(a)], one can observe that the spatial intensity distribution exhibits striking end confinements, as depicted in Fig. 3(a). In contrast, for the case of break mirror symmetry [the inset of Fig. 3(b)], it can be noticed that the representative boundary state begin to hybridize with the bulk continuums, manifested as a weakening of spatial confinements [Fig. 3(b)]. One can thus conclude that the in-bulk boundary states of the bilayer SSH-like model are robust against mirror-symmetry-preserving perturbation.

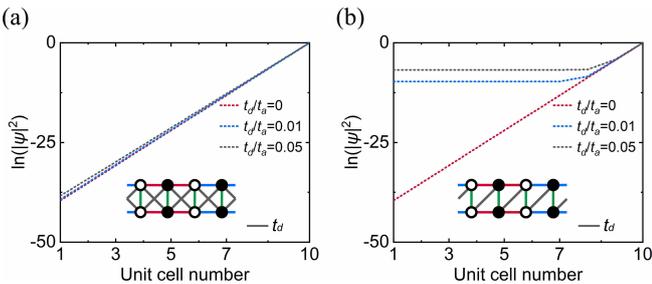

FIG. 3 (a)-(b) Sspatial intensity distribution for the configurations of the extra couplings is introduced in the bilayer SSH-like model that (a) preserves and (b) breaks mirror symmetry.

*Decorated SSH-like models.* In this section, we consider a decorated SSH-like model that includes ten sites per unit cell, as shown in Fig. 4(a), which is designed by adding additional sites in the bilayer SSH model [Fig. 4(b)]. Under the periodic boundary condition, the Hamiltonian of the decorated SSH-like model is given by

$$H_2 = \begin{pmatrix} O_{4\times 4} & \theta \\ \theta^\dagger & O_{6\times 6} \end{pmatrix}, \quad (5)$$

where $O_{n\times n}$ represents the $n \times n$ zero matrix, and $\theta$ is the $4 \times 6$ matrix

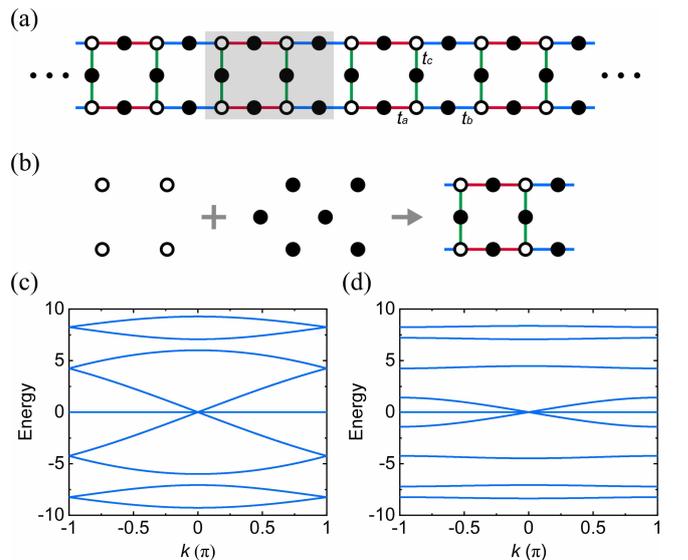

FIG. 4 (a) Schematic of the decorated SSH-like model. (b) The relation among the original model and decorated SSH-like model. (c)-(d) band structure with (c) $t_a = t_b = 3$, $t_c = 5$ and (d) $t_a = 1$, $t_b = 3$, $t_c = 5$.

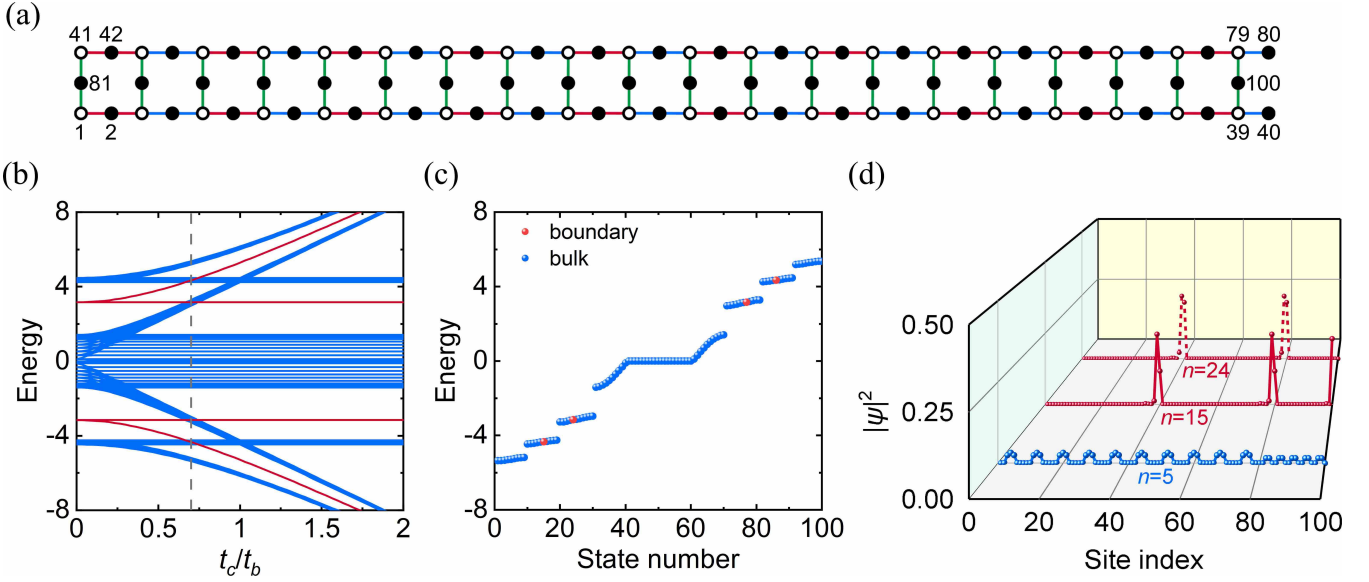

FIG. 5. (a) Schematic of the finite-size decorated SSH-like model. (b) Energy spectrum for the model in (a) with different $t_c$ but fixed $t_a = 1$, $t_b = 3$. (c) Energy spectrum at $t_c = 2.1$ [gray dotted line in (b)]. (d) The probability distributions of the bulk and boundary states for the decorated SSH-like model.

$$\theta = \begin{pmatrix} t_a & t_b e^{-ik} & t_c & 0 & 0 & 0 \\ t_a & t_b & 0 & t_c & 0 & 0 \\ 0 & 0 & t_c & 0 & t_a & t_b e^{-ik} \\ 0 & 0 & 0 & t_c & t_a & t_b \end{pmatrix}. \quad (6)$$

Where the hopping parameter $t_i$ ($i = a, b, c$) is shown in Fig. 4(a). It should be noticed that the Hamiltonian $H_2$ is chiral symmetric, satisfying $\gamma H_2 \gamma^{-1} = -H_2$, where $\gamma = \begin{pmatrix} I_{4 \times 4} & O_{4 \times 6} \\ O_{6 \times 4} & -I_{6 \times 6} \end{pmatrix}$. Due to the chiral symmetry, the squared Hamiltonian can be block diagonalized as follows

$$[H_2]^2 = \begin{pmatrix} h_{par} & O_{4 \times 6} \\ O_{6 \times 4} & h_{res} \end{pmatrix}, \quad (7)$$

where $h_{par} = \theta \theta^\dagger$ corresponds to the Hamiltonian of the bilayer SSH model with energy shift $t_a^2 + t_b^2 + t_c^2$, and $h_{res} = \theta^\dagger \theta$ is the residual Hamiltonian. Thus, the decorated SSH-like model can be viewed as the square-root version of the bilayer SSH model, i.e., correspond to the square-root topology, with the square-root relation reflected in Eq. (7). Although the inversion axis of this square-root system is offset with respect to the center of the unit cell, indicating the non-quantized topological invariant[13,28,29], the quantization can be recovered from the squared Hamiltonian, and the related original model has been studied in previous works[41,47,48]. The energy band structures for different coupling settings are depicted in Fig. 4(c)–(d). For the case of $t_a = t_b = 3$, $t_c = 5$ shown in Fig. 4(c), the 1st (3rd, 7th, 9th) bands touch the 2nd (4th, 8th, 10th) bands at $k = \pm\pi$. when $t_a < t_b$, these degeneracies are gapped out [Fig. 4(d)], providing the opportunity to realize the square-root topological states. Moreover, there are zero-energy flat bands (5th and 6th) that emerge in the form of two-fold degeneracy, with the number of these flat bands equal to the sublattice imbalance in the unit cell according to Lieb's theorem[49].

To examine the boundary states of the decorated SSH-like model, we consider a finite-size structure with 10 unit cells under OBC, as illustrated in Fig. 5(a). For the decorated SSH-like model, the hopping parameter $t_c$ does not affect the existence of the boundary states. To show it, we plot the eigenvalue as a function of $t_c / t_b$ while maintaining $t_a = 1$, $t_b = 3$ in Fig. 5(b). One can observe that the multiple boundary states persist when $t_c$ changes, and the four groups of boundary states come in two chiral pairs due to the chiral symmetry. To displays the existence of multiple in-bulk boundary states, the corresponding energy spectrum of the typical example at $t_c = 2.1$ is shown in the Fig. 5(c). Furthermore, the distributions of the representative boundary states and bulk states are shown in the Fig. 5(d), exhibiting that the boundary states are confined at the end region of the structure.

For the decorated SSH-like model, we can alternatively regard it as a combination of two typical square-root SSH models connected through the middle layer. For simplicity of comprehension, the Hamiltonian of the decorated SSH-like model for this scenario can be expressed as follows

$$H_3 = \begin{pmatrix} 0 & 0 & t_a & t_b e^{-ik} & t_c & 0 & 0 & 0 & 0 & 0 \\ 0 & 0 & t_a & t_b & 0 & t_c & 0 & 0 & 0 & 0 \\ t_a & t_a & 0 & 0 & 0 & 0 & 0 & 0 & 0 & 0 \\ t_b e^{ik} & t_b & 0 & 0 & 0 & 0 & 0 & 0 & 0 & 0 \\ t_c & 0 & 0 & 0 & 0 & 0 & t_c & 0 & 0 & 0 \\ 0 & t_c & 0 & 0 & 0 & 0 & 0 & t_c & 0 & 0 \\ 0 & 0 & 0 & 0 & t_c & 0 & 0 & 0 & t_a & t_b e^{-ik} \\ 0 & 0 & 0 & 0 & 0 & t_c & 0 & 0 & t_a & t_b \\ 0 & 0 & 0 & 0 & 0 & 0 & t_a & t_a & 0 & 0 \\ 0 & 0 & 0 & 0 & 0 & 0 & t_b e^{ik} & t_b & 0 & 0 \end{pmatrix}. \quad (8)$$

Here, the Hamiltonian $H_3$ can be considered to be constructed of three layers if the interlayer coupling $t_c$ is disregarded, in which two layers correspond to typical square-root SSH models. It is interesting to point out that this system satisfied the mirror symmetry with $M_{z,2} = \begin{pmatrix} O_{4\times4} & O_{4\times2} & I_{4\times4} \\ O_{2\times4} & I_{2\times2} & O_{2\times4} \\ I_{4\times4} & O_{4\times2} & O_{4\times4} \end{pmatrix}$. If we use the similarity transformation $U^{-1} M_{z,2} U = \tilde{M}_{z,2}$ with

$$U = \frac{1}{\sqrt{2}} \begin{pmatrix} I_{4\times4} & O_{4\times2} & -I_{4\times4} \\ O_{2\times4} & \sqrt{2} I_{2\times2} & O_{2\times4} \\ I_{4\times4} & O_{4\times2} & I_{4\times4} \end{pmatrix}, \quad (9)$$

the mirror operator is diagonalized into $\tilde{M}_{z,2} = \begin{pmatrix} I_{6\times6} & O_{6\times4} \\ O_{4\times6} & -I_{4\times4} \end{pmatrix}$. Meanwhile, $H_3$ can be divided into the subspaces after the similarity transformation with $U^{-1} H_3 U = \tilde{H}_3$, with $\tilde{H}_3$ is written as

$$\tilde{H}_3 = \begin{pmatrix} \lambda_1 & O_{6\times4} \\ O_{4\times6} & \lambda_2 \end{pmatrix}, \quad (10)$$

here, we introduced

$$\lambda_1 = \begin{pmatrix} O_{2\times2} & \eta_1 \\ \eta_1^\dagger & O_{4\times4} \end{pmatrix}, \quad (11)$$

$$\lambda_2 = \begin{pmatrix} O_{2\times2} & \eta_2 \\ \eta_2^\dagger & O_{2\times2} \end{pmatrix}, \quad (12)$$

with

$$\eta_1 = \begin{pmatrix} t_a & t_b e^{-ik} & \sqrt{2} t_c & 0 \\ t_a & t_b & 0 & \sqrt{2} t_c \end{pmatrix}, \quad (13)$$

$$\eta_2 = \begin{pmatrix} t_a & t_b e^{-ik} \\ t_a & t_b \end{pmatrix}, \quad (14)$$

where the subspace Hamiltonian $\lambda_1$ corresponds to the unconventional square-root version of the SSH chain, which has six sites per unit cell. The squared Hamiltonian of $\lambda_1$ is $\begin{pmatrix} \eta_1 \eta_1^\dagger & O_{2\times4} \\ O_{4\times2} & \eta_1^\dagger \eta_1 \end{pmatrix}$. Accordingly, $\eta_1 \eta_1^\dagger$ corresponds to the SSH chain with the on-site potential $t_a^2 + t_b^2 + 2 t_c^2$, and $\eta_1^\dagger \eta_1$ is the residual Hamiltonian. Moreover, the subspace Hamiltonian $\lambda_2$ corresponds to the conventional square-root version of the SSH chain.

As shown in Fig. 5(b)-(c), the boundary states of the decorated SSH-like model can be embedded in the bulk to form in-bulk boundary states. To investigate the robustness of the in-bulk boundary states, we introduce two types of weak perturbation that preserve and break mirror symmetry, respectively. The configuration that preserves mirror symmetry is shown in the inset of Fig. 6(a). In this case, the spatial intensity distribution in Fig. 6(a) shows the significant end confinements. In contrast, for the configuration that breaks mirror symmetry [Fig. 6(b)], the spatial intensity distribution is strongly influenced, indicating that

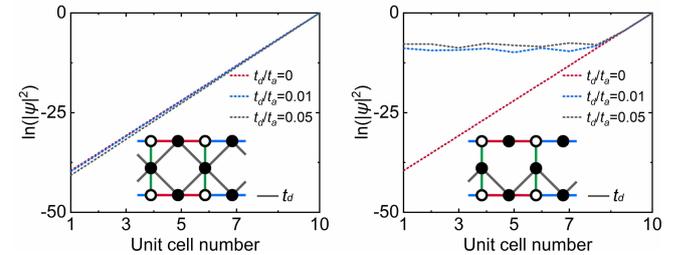

FIG. 6 (a)-(b) Sspatial intensity distribution for the decorated SSH-like model with extra couplings that (a) preserves and (b) breaks mirror symmetry.

hybridization occurred between the boundary and bulk states. These results indicate the in-bulk boundary states of the decorated SSH-like model are robust against mirror-symmetry-preserving perturbation.

*Proposal for circuit schemes.* Electric circuits have recently served as an appropriate platform for investigating topological physics due to the flexible designability of network structure[50-52]. Various exciting fields have been discussed by the platform of electric circuits, such as Chern insulators[53,54], topological Anderson insulators[55,56], and quadrupole insulators[57].

In this section, we propose circuit schemes to construct the bilayer SSH-like and decorated SSH-like models. The designed finite-size circuits composed of capacitors and inductors are shown in Fig. 7(a)–(b). Specifically, the capacitors $C_n$ ($n = a,b,c$) nconnected between nodes in our circuits represent the corresponding hoppings in the tight-binding method, and the grounding components of capacitors and identical inductors [Fig. 7(c)] are introduced to each node to maintain the same on-site potential.

The circuit response can be described by Kirchhoff's law $I = JV$, where circuit Laplacian $J$ relates the column vectors of input current $I$ and the corresponding voltage $V$. For bilayer SSH-like and decorated SSH-like circuits, the matrix form of the circuit Laplacian in momentum space can be expressed as

$$J_1 = -i\omega \begin{pmatrix} Q_1 & 0 & C_a & C_b e^{-ik} & C_c & \cdots & 0 \\ 0 & Q_1 & C_a & C_b & 0 & \cdots & 0 \\ C_a & C_a & Q_1 & 0 & 0 & \cdots & 0 \\ C_b e^{ik} & C_b & 0 & Q_1 & 0 & \cdots & C_c \\ C_c & 0 & 0 & 0 & Q_1 & \cdots & C_b e^{-ik} \\ \vdots & \vdots & \vdots & \vdots & \vdots & \ddots & \vdots \\ 0 & 0 & 0 & C_c & C_b e^{ik} & \cdots & Q_1 \end{pmatrix} \quad (15)$$

and

$$J_2 = -i\omega \begin{pmatrix} Q_2 & 0 & C_a & C_b e^{-ik} & C_c & \cdots & 0 \\ 0 & Q_2 & C_a & C_b & 0 & \cdots & 0 \\ C_a & C_a & Q_2 & 0 & 0 & \cdots & 0 \\ C_b e^{ik} & C_b & 0 & Q_2 & 0 & \cdots & 0 \\ C_c & 0 & 0 & 0 & Q_2 & \cdots & 0 \\ \vdots & \vdots & \vdots & \vdots & \vdots & \ddots & \vdots \\ 0 & 0 & 0 & 0 & 0 & \cdots & Q_2 \end{pmatrix}. \quad (16)$$

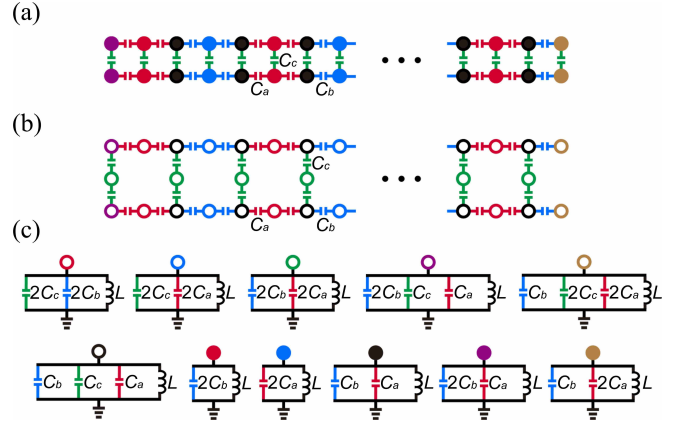

FIG. 7 (a)-(b) The circuit schemes of (a) bilayer SSH-like model and (b) decorated SSH-like model. (c) Corresponding grounding pattern.

with the matrix elements $Q_1 = -2C_a - 2C_b - C_c + 1/\omega^2 L$, $Q_2 = -2C_a - 2C_b - 2C_c + 1/\omega^2 L$. $J_1$ and $J_2$ are the circuit Laplacian of bilayer SSH-like and decorated SSH-like circuits, respectively. By utilizing the above schemes, one can readily investigate the topological phenomena of bilayer SSH-like and decorated SSH-like models in this work based on the platform of electric circuits.

*Conclusions.* In conclusion, we have investigated the multiple boundary states based on bilayer and decorated SSH-like models. The bilayer SSH-like model hosts multiple boundary states and allows for the coexistence of in-bulk and in-gap boundary states under OBC. Furthermore, we reveals the square-root topology based on the decorated SSH-like model, which host multiple in-bulk boundary states at appropriate hopping parameters. We also found the connections between this model and the square-root SSH models with four and six sites, respectively. Both the robustness of in-bulk boundary states within the above models are examined by introducing the extra hoppings. Moreover, the circuit schemes of the bilayer and decorated SSH-like models are presented to establish the foundation for further research into their intriguing phenomena based on classical systems. Our results not only provide a new perspective for studying multiple boundary states and square-root topology but also pave the way for future extensions to higher-dimensional systems.

*Acknowledgments.* The authors are grateful to the support from the Natural Science Fund of China under Grant No. 11774103, and Quanzhou City Science & Technology Program of China under Grand No. 2018C003.